%% file: SubfsGloryHologrammetry.tex
\documentclass[showpacs,preprintnumbers,amsmath,amssymb,twocolumn,prl]
{revtex4}

\usepackage{graphicx}% Include figure files
\usepackage{hyperref}

\begin{document}

\title{Subfemtosecond glory hologrammetry for vectorial optical waveform reconstruction}

\author{J. F. Tao$^1$}
%\email{jianfeitau@csrc.ac.cn}
\author{J. Cai$^2$}
\author{Q. Z. Xia$^3$}
\email{xia\_qinzhi@iapcm.ac.cn}
\author{J. Liu$^{4, 5}$}
\email{jliu@gscaep.ac.cn}

\affiliation{ $^1$Beijing Computational Science Research Center, Beijing 100193, China
}

\affiliation{$^2$School of Physics and Electronic Engineering, Jiangsu Normal University, Xuzhou 221116, China}

\affiliation{$^3$National Laboratory of Science and Technology on Computational Physics,
Institute of Applied Physics and Computational Mathematics, Beijing 100088, China}

\affiliation{$^4$Graduate School of China Academy of Engineering Physics, Beijing 100193, China}

\affiliation{$^5$CAPT, HEDPS, and IFSA Collaborative Innovation Center of MoE, Peking University, Beijing 100871, China}

\begin{abstract}

In this Letter, we propose a new method to characterize the temporal structure of arbitrary optical laser pulses with low pulse energies. This approach is based on strong field photoelectron holography with the glory rescattering effect as the underlying mechanism in the near-forward direction. Utilizing the subfemtosecond glory rescattering process as a fast temporal gate to sample the unknown light pulse, the time-dependent vectorial electric field can be retrieved from the streaking photoelectron momentum spectra. Our method avoids the challenging task of generation or manipulation of attosecond pulses and signifies important progress in arbitrary optical waveform characterization.

\end{abstract}

\maketitle

Probing or manipulation of ultrafast electron dynamics on a subfemtosecond($\le10^{-15}~s$) or attosecond($\sim 10^{-18}~s$) timescale necessitates ultrashort laser pulses lasting only a few or near-single optical cycles with controllable waveforms\cite{Goulielmakis769, PhysRevLett.103.213003, Schiffrin2012, Ishii2014, Garg2016, Sommer2016, PhysRevX.9.031004}. Developments in frequency comb technology combined with pulse-shaping methods have allowed arbitrary electromagnetic waveforms to be synthesized at optical frequencies\cite{Del'Haye2007, Jiang2007, Cundiff2010, Chan1165, Wirth195, Schliesser2012}. Knowledge of the temporal structure of these light pulses is a prerequisite for subsequent applications. Traditional characterization techniques, such as frequency-resolved optical gating(FROG), spectral phase interferometry for direct electric field reconstruction(SPIDER) or dispersion scan(d-scan), have been used to measure the spectral/temporal amplitude/phase or dispersion/chirp of short pulses\cite{doi:10.1063/1.1148286, Iaconis:98, Miranda:12}. However, the phase-matching problem of nonlinear crystals and the deficiency in determining the absolute phase(carrier-envelope phase, CEP) both limit their applicability. Instead, direct access to the time-domain electric field $\textbf{E}_L(t)$ requires a fast nonlinear response that is significantly shorter than an optical cycle\cite{Wyatt:16, Park:18}.

 Advancements in strong field physics have provided such ultrashort temporal gates. One widely used technique is attosecond streak camera\cite{Sansone443, Goulielmakis1267, PhysRevLett.88.173903, Boge:14}: isolated attosecond extreme ultraviolet(XUV) pulses generated by higher-order harmonic generation(HHG) processes are used to ionize atoms\cite{PhysRevA.49.2117, Baltuska2003, Witting_2012, ABEL20099, Ferrari2010, Sola2006}. The ejected photoelectrons are then streaked to different final energies by the test laser field whose waveform is to be measured. The temporal structure of both the test laser and the attosecond XUV pulse can be accurately reconstructed from the streaking photoelectron spectra\cite{PhysRevA.71.011401, Lucchini:15}. Two other all-optical characterization methods, petahertz optical oscilloscope and attosecond spatial interferometry, both utilize the subfemtosecond tunneling-recombination process during HHG generation as the temporal gate to sample the test optical laser field\cite{Kim2013, Carpeggiani2017}.

 Although these recent characterization techniques yield good performance, their requirement of generation or manipulation of broadband isolated attosecond XUV pulses is still very challenging to meet\cite{Wang_2009, Krausz2014, Chini2014}. In this Letter, we propose a new method to extract the waveforms of unknown laser pulses with commonly used strong near-infrared(NIR) table-top laser light as a pump field to irradiate the atoms. Our proposal utilizes facilities from the strong field ionization and strong field photoelectron holography(SFPH) fields\cite{Huismans61}, and information of the weak test laser pulses is imprinted in the holographic interference fringes of the final photoelectron momentum distribution(PMD).

%\section{theory}
A strong NIR laser is able to tunnel ionize atoms, and the liberated photoelectron may be driven back and elastically scatter off the parent ion at a later time\cite{PhysRevLett.71.1994}. Concerning SFPH, strong field tunneling ionization plays the role of an atomic-level beamsplitter: after tunneling, part of the photoelectron wavepacket less impacted by the ionic Coulomb potential forms a reference wave. The other part, termed the signal wave, is steered around and scatters off the atomic core. The hologram stemming from interference of the reference and signal waves at the detector encodes spatiotemporal information about the interaction of the electron-ion system. Recently, the interpretation of SFPH  has been improved by the discovery of the glory rescattering effect in strong field ionization\cite{PhysRevLett.121.143201}.

\begin{figure}
    \includegraphics[width=0.9\linewidth]{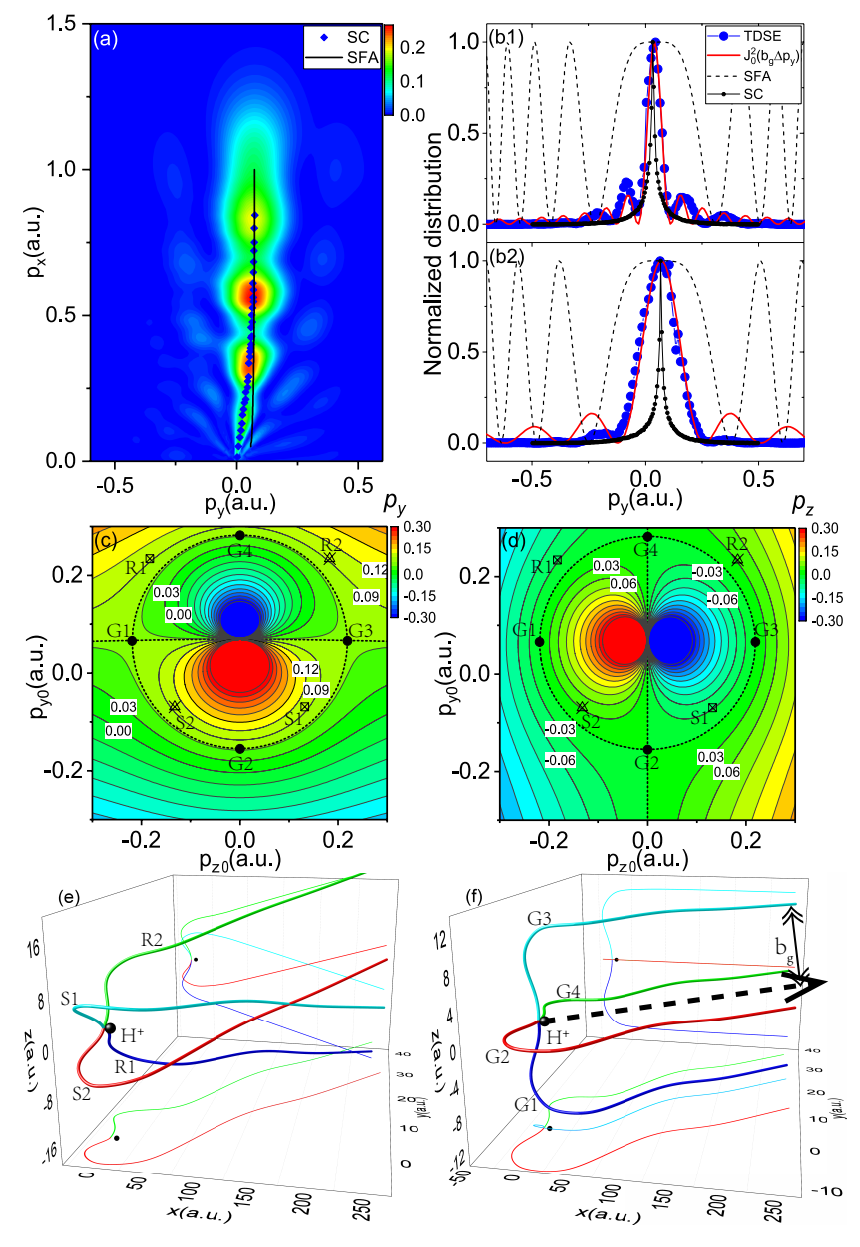}
    \caption{\label{fig:PMD}(a) PMD calculated using the TDSE in the polarization plane($p_z = 0$), and ionization of H atoms by an OTC laser field. The glory interference maxima(GIM) estimated by the semiclassical method(SC, blue diamonds) and strong field approximation(SFA, black solid line) are also presented. (b1-b2) Normalized transverse momentum distribution with $p_x = $ 0.2(b1) and 0.6(b2). Note that the SFA curves have been shifted to match the peak positions calculated by the SC and TDSE. (c)(d) Contour plots of the deflection functions with $\eta_0=0.3$ and $p_x \approx 0.6$. The circle contour indicates the initial conditions of glory trajectories(GTs), while four typical GTs(G1, G2, G3 and G4) are shown in (f) with XY and XZ projections. (e)Two pairs of signal/reference trajectories with initial conditions from regions (S1, R1) and (S2, R2). See the text for more details.  }
\end{figure}

For theoretical demonstration purposes, a fundamental pump laser field with a wavelength of $800nm$ and an intensity of $1.5 \times 10^{14} W/cm^2$ is used to ionize hydrogen atoms: $\label{eqn:pumpLaser}
\textbf{E}_0(t) = \epsilon_0 \cos^2(\frac{\pi t}{T_0}) \cos(\omega_0 t) \mathbf{\hat{x}}$, where $T_0 = 3\times \frac{2\pi}{\omega_0}$, with the time duration only three optical cycles to eliminate multiple rescattering effects.  Fig.~\ref{fig:PMD}(a) illustrates the PMD in the polarization plane simulated using the time-dependent Schr\"{o}dinger equation(TDSE), with an orthogonally polarized two-color(OTC) laser  field. The test  laser pulse has a wavelength of $1600nm$, an intensity of $2.4\times 10^{11}W/cm^2$, and a time duration of four optical cycles: $\textbf{E}_L(t) = \epsilon_L \cos^2(\frac{\pi t}{T_L}) \cos (\omega_L t) \mathbf{\hat{y}}$, with $T_L = 4\times \frac{2\pi} {\omega_L}$. The spider-like interference fringes characteristic of SFPH are clearly visible\cite{Huismans61}. Unless stated otherwise, atomic units will be used throughout.

Without considering the Coulomb potential, the phase difference responsible for the hologram between the signal and reference photoelectron waves can be derived using strong field approximation(SFA) or approximations from the path integral method as follows\cite{Huismans61, PhysRevLett.122.183202, PhysRevA.100.023413}:
\begin{eqnarray}
\label{eqn:modified_phase_shift}
\delta \phi \approx \frac{1}{2} (\textbf{p}_\perp-\textbf{k}_L)^2 (t_r - t_0^{ref})
\end{eqnarray}
in which $\textbf{p}_\perp$ is the asymptotic photoelectron momentum perpendicular to the fundamental laser polarization. $t_r$ is the rescattering time, and $t_0^{ref}$ is the ionization time of the reference photoelectron wave. The intermediate canonical momentum between tunneling and rescattering is $\textbf{k}_L = -\frac{1}{t_r - t_0^R} \int^{t_r}_{t_0^R}\textbf{A}_L(t^\prime)dt^\prime$ to ensure that the electron travels back to the ion, while $t_0^R$ is the ionization time of the rescattering wavepacket. Generally, for near-forward rescattering with a small transverse momentum $\textbf{p}_\perp$, the tunneling time for reference and rescattering(signal) quantum paths are approximately the same: $t_0^{ref} \approx t_0^R$. $\textbf{A}_L(t) = -\int^t\textbf{E}_L(t^\prime)dt^\prime$ is the vector potential of the weak test laser field.

However, a $\cos(\mathcal{R}e(\delta \phi))$-like peak structure derived from Eqn.~\ref{eqn:modified_phase_shift} for the transverse momentum distribution(black dashed lines in Fig.~\ref{fig:PMD}(b1)(b2) for different asymptotic longitudinal momenta $p_x = 0.2, 0.6$)  fails to reproduce the TDSE results(blue dotted lines). This problem can be clarified from the semiclassical(SC) perspective of the Feynman path integral method, which dictates that the dominant contributions come from the regions around the classical trajectories. Fig.~\ref{fig:PMD}(c)(d) depict the contour plots of the deflection functions $\textbf{p}_\perp = \textbf{p}_\perp(\eta_0, \textbf{p}_{\perp 0})$ obtained by solving Newton's equation of motion after the electron emerges at $\eta_0=\omega_0t_0 = \omega_0\mathcal{R}e(t_0^{ref}) \approx 0.3$\cite{liujie2014, PhysRevA.97.063409}. Due to Coulomb potential influence, the $p_{y0}\_p_{z0}$ plane can be divided into four signal/reference regional pairs: (S1, R1), (S2, R2), (S3, R3) and (S4, R4)(with the latter two not shown). For the final photoelectron momentum originating from inside these pairs, only two classical trajectories are found(Fig.~\ref{fig:PMD}(e)); however, infinite classical trajectories stemming from the circle contour dividing the signal and reference regions all contribute to the same asymptotic momentum(Fig.~\ref{fig:PMD}(f) depicts four such classical orbits).

This phenomenon is analogous to the (forward) glory effect in quantum scattering theory\cite{FORD1959259}. The contributions of infinite so-called glory trajectories(GTs) to the final momentum distribution should be summed up. Referring to Eqn.~\ref{eqn:modified_phase_shift}, for simplicity, consider the case with only the NIR fundamental pulse; for a small deviation $\Delta p_\perp$ from the forward direction, we have $\Delta (\mathcal{R}e(\delta \phi)) \sim \Delta p_\perp p_{\perp 0}(t_r - t_0)\sim \Delta p_\perp b_g$, where $p_{\perp 0} \ne 0$ is the initial transverse momentum with the Coulomb potential involved. $b_g \sim p_{\perp 0}(t_r - t_0)$ is interpreted as the asymptotic impact factor of GTs(Fig.~\ref{fig:PMD}(f))\cite{PhysRevLett.121.143201}. Then, the transverse momentum distribution  in the near-forward direction is $\label{eqn:phase_shift_glory}
f(p_\perp) \propto | \frac{1}{2\pi}\int_0^{2\pi} e^{i \Delta p_\perp b_g \cos\theta} d\theta|^2 =J_0^2(b_g\Delta p_\perp)$. In an OTC field, this would result in $f(p_\perp) \propto J_0^2(b_g|\textbf{p}_\perp-\textbf{p}_L|)$. $\textbf{p}_L$ is the transverse momenta corresponding to the primary glory interference maxima(GIM)(on the circle contour in Fig.~\ref{fig:PMD}(c)(d), $|\textbf{p}_\perp-\textbf{p}_L| \equiv 0$). This result has successfully interpreted the near-forward SFPH interference fringes in PMD\cite{PhysRevLett.121.143201, PhysRevA.100.023413, PhysRevA.100.023419}.

Using this SC photoelectron trajectory method, $b_g$ can be retrieved by back-propagation for each $p_x$\cite{PhysRevLett.121.143201}. The resulting squared-Bessel-like peak structure(red solid lines in Fig.~\ref{fig:PMD}(b1)(b2)) agrees very well with the TDSE simulation. A SC trajectory Monte Carlo simulation also reproduces the position of the GIM $\textbf{p}_L$(blue diamonds in Fig.~\ref{fig:PMD}(a), black dotted lines in Fig.~\ref{fig:PMD}(b1)(b2)). An approximation of this position can be found from Eqn.~\ref{eqn:modified_phase_shift}\cite{PhysRevA.95.011402}:  $\textbf{p}_L \approx \mathcal{R}e(\textbf{k}_L)$. This result, shown in Fig.~\ref{fig:PMD}(a)(black solid line), describes the TDSE/SC simulations quite well, especially for larger longitudinal photoelectron momentum $p_x$. The deviation for smaller $p_x$ is due to the Coulomb effects.

\begin{figure}
    \includegraphics[width=0.8\linewidth]{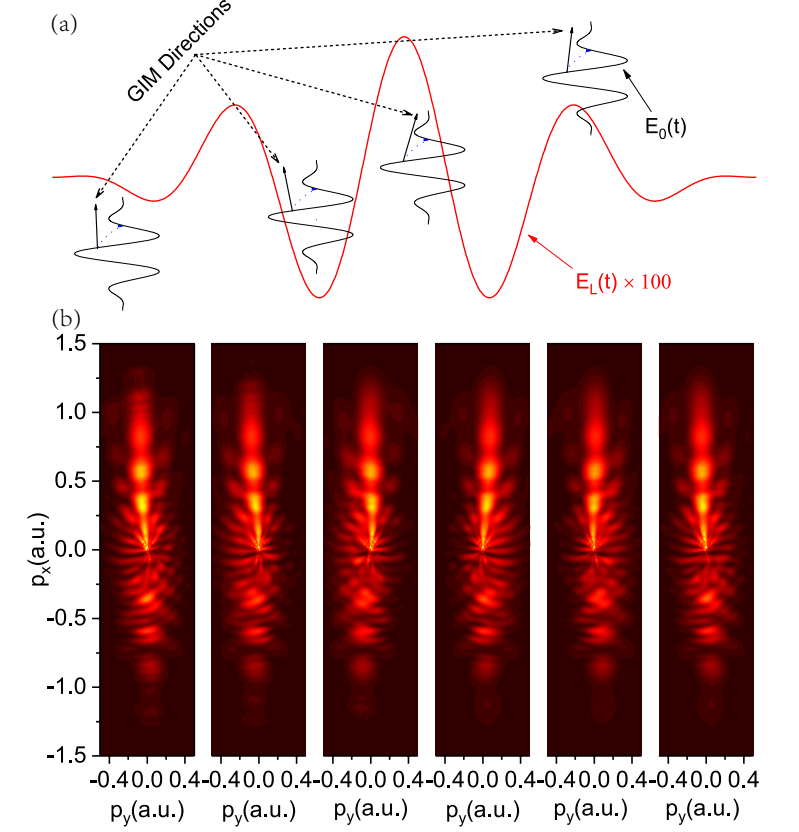}
    \caption{\label{fig:PMD_for_different_timedelay}(a) Illustration of the sampling of a test laser field with the subfemtosecond glory rescattering process. Blue dashed arrows indicate the subcycle excursion of the tunneled electrons. (b) Integrated PMD simulated using the TDSE with time delay $\Delta \tau = \frac{n}{6}~o.c.$, where $n = -6, -4, -2, 2, 4, 6$(left to right). $1~o.c. = \frac{2\pi}{\omega_0}$.}
\end{figure}

Therefore, adding a weak test laser $\textbf{E}_L$($\perp \textbf{E}_0$) introduces an extra factor into the phase difference between the reference and signal photoelectron waves(Eqn.~\ref{eqn:modified_phase_shift}) or, classically, slightly perturbs the whole bunch of glory rescattering trajectories(Fig.~\ref{fig:PMD}(f)). One of the consequences is a peak shift of the asymptotic transverse momentum distribution, the same as that in nondipole strong field ionization \cite{PhysRevA.95.011402, PhysRevLett.113.243001, PhysRevA.98.023412, PhysRevA.97.063409, PhysRevLett.118.163203, Hartung2019}. Therefore, we can utilize the subfemtosecond glory rescattering process as a fast temporal gate to sample a test laser pulse by varying the time delay between the fundamental and weak light pulses(Fig.~\ref{fig:PMD_for_different_timedelay}(a)):
\begin{eqnarray}
\label{eqn:peak_position_SFA}
% \nonumber % Remove numbering (before each equation)
  \textbf{p}_L(\Delta \tau) \approx \mathcal{R}e\{ -\frac{1}{t_r - t_0^R} \int^{t_r+ \Delta \tau}_{t_0^R+ \Delta \tau}\textbf{A}_L(t )dt \}
\end{eqnarray}
Fig.~\ref{fig:PMD_for_different_timedelay}(b) illustrates the TDSE simulation of PMDs with different time delays; the GIM oscillate with $\Delta \tau$.
If the test light pulse does not contain frequency components that are larger than about $\omega_0$, we can derive the approximate waveform of the test light from the measured GIM as $\textbf{A}_L(t) \approx - \textbf{p}_L(t-t_\alpha) + t^2_\beta \frac{d^2}{dt^2}\textbf{p}_L(t-t_\alpha)$ for larger longitudinal momentum($p_x \sim (50 \% - 90 \%)\times \frac{\epsilon_0}{\omega_0}$)\cite{PhysRevA.100.043410}, the second term on the right hand side is much smaller than the first in the present setup. $t_{\alpha, \beta}$ are small time parameters determined by the fundamental ionizing laser field(See the Supplementary for more details). Current experiments can measure the smallest transverse momentum amounting to that carried by a few photons, which is on the order of $\delta p_c \sim 10^{-3}~a.u.$\cite{PhysRevLett.106.193002, PhysRevLett.113.243001, Hartung2019}. It is sufficient to resolve the peak positions in our scheme($\frac{\epsilon_L}{\omega_L} \gg \delta p_c$). In the following demonstrations we have also chosen the upper bound of the difference between two consecutively sampled peak shifts, estimated as $\delta p_\perp  \sim \epsilon_L \frac{\sinh\omega_L t_i}{\omega_L t_i} \delta t$, to be slightly larger: $\delta p_\perp \gtrsim \delta p_c$, where $t_i = \mathcal{I}m(t_0^R)$ and $\delta t$ is the time-step associated with changing the time delay between the fundamental and test laser fields.

 %Based on Eqn.~\ref{eqn:peak_position_SFA}, an upper bound for the momentum $\textbf{p}_L$ increment can be estimated as $\delta p_\perp  \sim \epsilon_L \frac{\sinh\omega_L t_i}{\omega_L t_i} \delta t$, where $t_i = \mathcal{I}m(t_0^R)$ and $\delta t$ is the time-step associated with changing the time delay between the fundamental and test laser fields. In the following demonstrations we have chosen this value to be larger than the smallest transverse momentum measurable in experiments, amounting to that carried by a few photons, which is on the order of $10^{-3}~a.u.$\cite{PhysRevLett.106.193002, PhysRevLett.113.243001, Hartung2019}.

%In the following simulations, we have chosen laser parameters and timestep to satisfy this restriction.

\begin{figure}
\includegraphics[width=0.9\linewidth]{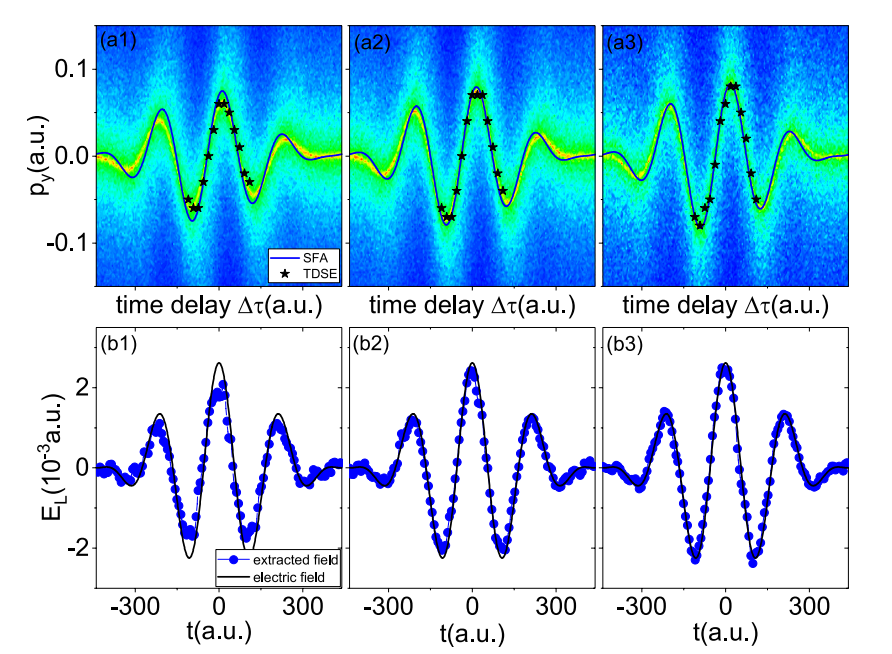}
\caption{\label{fig:steaking_for_different_px} (a1-a3) Streaking spectra of the photoelectron momentum along the test laser polarization direction versus time delay for $p_x =$ 0.4(a1), 0.6(a2) and 0.8(a3). The TDSE results(black stars) well fit the SC trajectory Monte Carlo simulation results. Blue solid lines are estimated using Eqn.~\ref{eqn:peak_position_SFA}. (b1-b3) Corresponding electric fields(blue dotted lines) extracted from the GIM compared with the actual field(black solid lines).}
\end{figure}

The streaking photoelectron spectra of the transverse momentum distribution versus time delay are presented in Fig.~\ref{fig:steaking_for_different_px}(a1)(a2)(a3) for different final longitudinal momenta $p_x = 0.4, 0.6, 0.8$. The same test laser light is used as in Fig.~\ref{fig:PMD}. TDSE results(black stars) agree very well with the SC simulation results. The SFA calculation yields a good approximation(blue solid lines in Fig.~\ref{fig:steaking_for_different_px}(a1)(a2)(a3)). More precisely, the electric field of the test laser pulse can be directly solved from Eqn.~\ref{eqn:peak_position_SFA}:
\begin{eqnarray}
\textbf{E}_L(t) = \frac{1}{i\pi}\int_{-\infty}^\infty \frac{\omega \tilde{\textbf{p}}_L(\omega)}{a(\omega)+a^\ast(-\omega)} e^{i \omega t}d\omega
\end{eqnarray}
$\tilde{\textbf{p}}_L(\omega) = \int_{-\infty}^\infty \textbf{p}_L(\Delta \tau)e^{-i \omega \Delta \tau}d\Delta \tau$ is the corresponding Fourier transform, and $a(\omega) = -\int_{t_0^R}^{t_r}e^{i \omega t^\prime}dt^\prime/(t_r-t_0^R)$. The extracted electric field is depicted as blue dotted lines in Fig.~\ref{fig:steaking_for_different_px}(b1)(b2)(b3). It reproduces the original test laser electric field(black solid lines).

\begin{figure}
\includegraphics[width=0.9\linewidth]{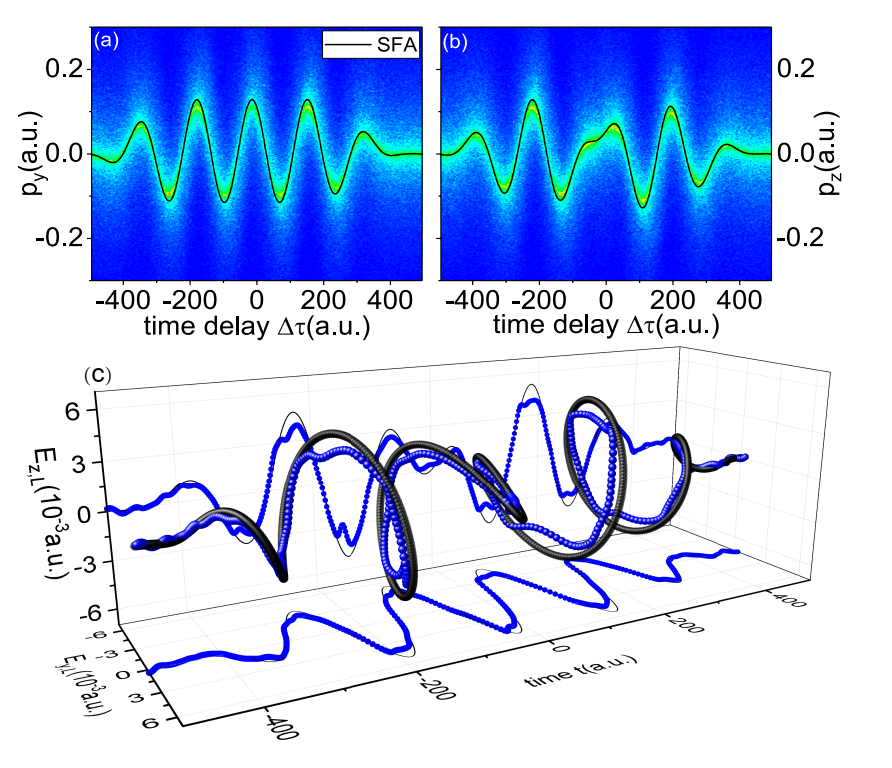}
\caption{\label{fig:moreWaves} (a-b) Streaking photoelectron momentum spectra for two independent polarization directions of the synthesized test laser light with time-varying ellipticity($p_x = 0.8$). (c) 3D representation of the extracted electric field(blue spheres). The result is compared to the synthesized waveform(black spheres); the reconstructed electric fields in the $\mathbf{\hat{y}}$ and $\mathbf{\hat{z}}$ directions(blue dotted lines) are also shown in the projections, alongside the respective actual fields(black solid lines).     }
\end{figure}

The test optical laser is superposed on the fundamental pump pulse with perpendicular polarization, and in principle, the waveform of a test laser pulse with complex polarization states can be measured and reconstructed\cite{Carpeggiani2017}. As an example, the streaking photoelectron spectra in two independent polarization directions are shown in Fig.~\ref{fig:moreWaves}(a)(b) for a light pulse with time-varying ellipticity synthesized by two counter-propagating circularly polarized laser beams: $\textbf{E}_l = \epsilon_L f_L(t-t_d/2)(\cos(\omega_L t+\frac{\pi}{4})\mathbf{\hat{y}}+\sin(\omega_L t+\frac{\pi}{4})\mathbf{\hat{z}})$ and $\textbf{E}_r = \epsilon_L f_L(t+t_d/2)(\cos(\omega_L t+\frac{\pi}{4})\mathbf{\hat{y}}-\sin(\omega_L t+\frac{\pi}{4})\mathbf{\hat{z}})$\footnote{The carrier frequency is $\omega_L = 0.038 ~a.u.$, corresponding to a wavelength of 1200nm, with a time duration of $T_L = 4\times \frac{2\pi}{\omega_L}, T_d = T_L/2$. And $\epsilon_L = 0.08 \times \epsilon_0$.}.  The same retrieval algorithm is used to simultaneously extract the two electric fields(Fig.~\ref{fig:moreWaves}(c)). For all of these complex test light conditions, our method yields good results. In the frequency domain, we have the relationship $\tilde{\textbf{p}}_L(\omega) = r(\omega) \tilde{\textbf{A}}_L(\omega)$, the amplitude of the frequency response function $r(\omega)$ is approximately unity until up to about $\omega_0$,  so although for demonstration purposes we have mostly used near-monochromatic pulses, this approach is also suitable for retrieval of optical waveforms with broad spectral bandwidths. By decreasing the wavelength of the fundamental ionizing laser field to, e.g. $400nm$, this method can be used to measure the electromagnetic waveforms in the visible, infrared and even terahertz regimes. Moreover, even though the proposed procedure requires that the time delay $\Delta \tau$ be continuously varied, single-shot measurement may be achieved by distributing the atoms spatially and using the spatial dependence of the propagating electromagnetic wave $A(\omega t- \textbf{k}\cdot\textbf{r})$ to provide the time delay, where $k = \omega /c$ is the wave vector.

In conclusion, by leveraging the subfemtosecond Coulomb glory rescattering effect as a fast temporal gate, we can sample arbitrary optical waveforms directly in the time domain with electron spectroscopy and reconstruct the temporal structure of the vectorial optical laser pulses. Our method completely avoids the use of attosecond XUV optics, and a conventional experimental setup related to strong field ionization research is sufficient to provide the required data. Our results will facilitate the study of ultrafast electron dynamics in attosecond physics.

%J. Cai contributes significantly to the TDSE simulations.

%\section*{Acknowledgement}
\begin{acknowledgments}
J. Cai contributed significantly to the time-dependent Schr\"{o}dinger equation simulations.

This work is supported by the National Natural
Science Foundation of China (Grants No. 11674034,
No. 11775030, No. 11974057 and No. 11447015) and
NSAF (Grants No. U1930402 and No. U1930403).
\end{acknowledgments}

\bibliographystyle{apsrev4-1}
%\bibliography{references}
\input{SubfsGloryHologrammetry.bbl}

\end{document}

\title{Supplemental material for `Subfemtosecond glory hologrammetry for vectorial optical waveform reconstruction'}

\author{J. F. Tao}
%\email{jianfeitau@csrc.ac.cn}
\author{J. Cai}
\author{Q. Z. Xia}
\email{xia\_qinzhi@iapcm.ac.cn}
\author{J. Liu}
\email{jliu@gscaep.ac.cn}
%\author{J. F. Tao$^{1 \dagger}$, J. Cai$^2$, Q. Z. Xia$^3$, J. Liu$^{3, 4}$}

%\pacs{05.45.-a, 05.60.Gg, 37.10.Jk}
\maketitle

%\section{Numerical simulation with time-dependent Schr\"{o}dinger equation(TDSE)}
\section{Strong field photoelectron holography as the interference of reference and signal waves}
Thanks to the pioneering work on strong field photoelectron holography(SFPH)\cite{Huismans61}, the interference pattern in the final momentum distribution can be attributed to the interference of electron waves which reach the detector directly with little Coulomb disturbance(reference or direct wave) and that which scatters off the parent ion(signal or rescattering wave). The phase difference between the reference and signal waves mostly determines the interference pattern in the momentum distribution:
\begin{eqnarray}
P = |M_s + M_r|^2
 =  |M_s|^2+|M_r|^2+2|M_s||M_r|\cos(\Delta S)
\end{eqnarray}

From the derivation by SFA, the phase factors for the reference and signal waves can be approximated respectively as:
\begin{eqnarray}
\label{eqn:phaseD}
S_{ref} = \frac{1}{2}\int_{t_0^{ref}}^\infty d\tau (\textbf{p}+\textbf{A}(\tau))^2 - I_p t_0^{ref}
\end{eqnarray}
and
\begin{eqnarray}
\label{eqn:phaseR}
S_{signal} = \frac{1}{2}\int_{t_r}^\infty d\tau (\textbf{p}+\textbf{A}(\tau))^2 + \frac{1}{2}\int_{t_0^R}^{t_r} d\tau(\textbf{k}+\textbf{A}(\tau))^2
- I_p t_0^R
\end{eqnarray}
$I_p$ is the ionization potential of the atom. $\textbf{k}$ is the intermediate canonical momentum between tunneling and rescattering for the signal wave. $\textbf{p}$ is the final asymptotic photoelectron momentum.

%\begin{figure}
%\includegraphics[bb=0bp 0bp 851bp 404bp,clip,scale=0.45]{figure1}
    %\includegraphics[width=0.6\linewidth]{img/TDSE_SFA.pdf}
    %\caption{\label{fig:PMD_pz=0} Strong field photoelectron holography in laser polarization plane($p_z = 0$) with an OTC field calculated by %TDSE(left) and SFA(right). Black solid line in the right panel is the interference maxima estimated by SFA.}
%\end{figure}

To solve for the various tunneling and rescattering time($t_0^{ref}, t_0^R, t_r$), saddle point approximation will be used for the reference(Eqn.~\ref{eqn:phaseD}) and signal(Eqn.~\ref{eqn:phaseR}) wave respectively:
\begin{eqnarray}
\frac{1}{2}(\textbf{p}+\textbf{A}(t_0^{ref}))^2 + I_p = 0
\end{eqnarray}
and
\begin{eqnarray}
\label{eqn:saddle_point_equation}
\frac{1}{2}(\textbf{k}+\textbf{A}(t_0^R))^2 + I_p  =  0  \nonumber \\
(\textbf{k}+\textbf{A}(t_r))^2  =  (\textbf{p}+\textbf{A}(t_r))^2 \nonumber \\
\int_{t_0^R}^{t_r}(\textbf{k}+\textbf{A}(\tau)) d\tau  =  0
\end{eqnarray}

The phase difference can be derived as:
\begin{eqnarray}
\label{eqn:phaseDifference}
\Delta S = \frac{1}{2}\int_{t_0^{ref}}^{t_r} d\tau (\textbf{p}+\textbf{A}(\tau))^2 - \frac{1}{2}\int_{t_0^R}^{t_r} d\tau(\textbf{k}+\textbf{A}(\tau))^2
 +  I_p(t_0^R-t_0^{ref})
\end{eqnarray}

The test weak laser field has negligible influence on the derivation of the tunneling and rescattering time $t_0^{ref}, t_0^R, t_r$.
Moreover, for forward scattering with small transverse momentum $\textbf{p}_\perp$, analysis shows that\cite{PhysRevLett.121.253203}: $t_0^R \approx t_0^{ref}, k_x \approx p_x$ and $Im(t_r) \approx 0$, with these simplifications, the phase difference in Eqn.~\ref{eqn:phaseDifference} for near-forward scattering can be derived as:
\begin{eqnarray}
\Delta S &\approx& \int_{t_0^{ref}}^{t_0^R}(\frac{1}{2}(p_x+A_x(\tau))^2+I_p) d\tau
 + \frac{1}{2}\int_{t_0^{ref}}^{t_r}((\textbf{p}_\perp+\textbf{A}_\perp(\tau))^2-(\textbf{k}_\perp+\textbf{A}_\perp(\tau))^2)d\tau \nonumber \\
&=& \frac{1}{2}\int_{t_0^{ref}}^{t_r}(\textbf{p}_\perp^2-\textbf{k}_\perp^2+2\textbf{p}_\perp \cdot \textbf{A}_\perp(\tau)-2\textbf{k}_\perp \cdot \textbf{A}_\perp(\tau))d\tau \nonumber \\
&=& \frac{1}{2}(\textbf{p}_\perp^2-\textbf{k}_\perp^2)(t_r-t_0^{ref})+(\textbf{p}_\perp-\textbf{k}_\perp) \cdot \int_{t_0^{ref}}^{t_r}\textbf{A}_\perp(\tau)d\tau \nonumber \\
&=& \frac{1}{2}(\textbf{p}_\perp^2-\textbf{k}_\perp^2)(t_r-t_0^{ref})+(\textbf{p}_\perp-\textbf{k}_\perp) \cdot (-\textbf{k}_\perp)(t_r-t_0^{ref}) \nonumber \\
&=& \frac{1}{2}(\textbf{p}_\perp-\textbf{k}_\perp)^2(t_r - t_0^{ref}) \nonumber \\
&=& \frac{1}{2} ((p_y - k_y)^2+(p_z-k_z)^2) (t_r - t_0^{ref})
\end{eqnarray}
in which
\begin{eqnarray}
\label{eqn:intermediateK}
k_{y(z)} = -\frac{1}{t_r - t_0^R} \int^{t_r}_{t_0^R}A_{y(z)}(t^\prime)dt^\prime
\end{eqnarray}
In the above derivation, we have used the relationship Eqn.~\ref{eqn:saddle_point_equation}.
%The SFA-estimated SFPH and the interference maxima are shown in Fig.~\ref{fig:PMD_pz=0} right panel, with the TDSE result in left panel. For small %longitudinal momenta, SFA overestimates the maximum shift due to negligence of Coulomb effects.

\begin{figure}
    \includegraphics[width=0.55\linewidth]{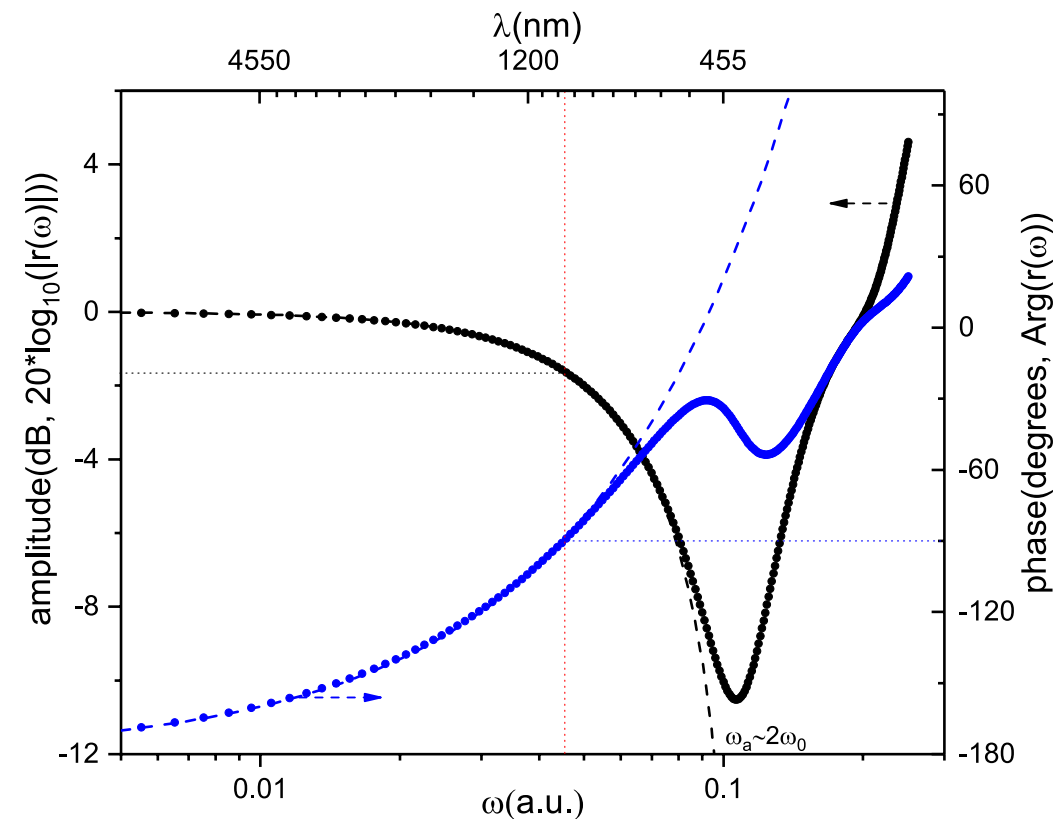}
    \caption{\label{fig:FRF}Amplitude(black dots) and phase(blue dots) of the frequency response function. Black and blue dashed lines represent the approximate FRF: $r(\omega) \approx (1- t^2_\beta \omega^2)e^{i(-\pi+t_\alpha \omega)}$.}
\end{figure}

To retrieve the test laser waveform  from the streaked photoelectron momentum distribution, experimentally extracting the peak shift of the transverse momentum distribution $f(p_\perp)$ is needed. Denoting this peak shift for every time delay which corresponds to the GIM as $\textbf{p}_L(\Delta \tau)$, then we have the approximation from Eqn.~\ref{eqn:intermediateK}:
\begin{eqnarray}
\label{eqn:peak_shift_SFA}
\textbf{p}_L(\Delta \tau)& \approx &\mathcal{R}e\{ -\frac{1}{t_r - t_0^R} \int^{t_r+ \Delta \tau}_{t_0^R+ \Delta \tau}\textbf{A}_L(t )dt \}\nonumber \\
& = & \mathcal{R}e\{ -\frac{1}{t_r - t_0^R} \int^{t_r}_{t_0^R}\textbf{A}_L(t+\Delta \tau)dt\}
\end{eqnarray}
Take the Fourier transform with variable $\Delta \tau$ of both sides, we have:
\begin{eqnarray}
\label{eqn:frequencyDomain}
\tilde{\textbf{p}}_L(\omega) = \tilde{\textbf{A}}_L(\omega) \frac{a(\omega)+a^\ast(-\omega)}{2}
\end{eqnarray}
$\tilde{\textbf{A}}_L(\omega) = \int_{-\infty}^\infty \textbf{A}_L(t)e^{-i \omega t}dt$, $\tilde{\textbf{p}}_L(\omega) = \int_{-\infty}^\infty \textbf{p}_L(\Delta \tau)e^{-i \omega \Delta \tau}d\Delta \tau$ are the corresponding Fourier transforms, $a(\omega) = -\int_{t_0^R}^{t_r}e^{i \omega t^\prime}dt^\prime/(t_r-t_0^R)$. Defining the frequency response function(FRF) of our measurement process $r(\omega) = \frac{a(\omega)+a^\ast(-\omega)}{2}$, its amplitude and phase are plotted in Fig.~\ref{fig:FRF}. We can clearly see that no frequency components contained in the test pulse $\textbf{A}_L(t)$ are significantly suppressed($|r(\omega)|\ge 0.3$, for when $p_x=0.8$ with the current fundamental ionizing field), so there is no strict restriction on the frequency bandwidth of the test pulse from this perspective. Actually we do not intend our method to measure electromagnetic waves with very short wavelengths. Other restrictions will be simply discussed below.

Because we use Fast Fourier Transform(FFT) algorithm in the retrieval process, the timestep $\delta t$ of the delay $\Delta \tau$ should not be chosen arbitrarily. Since $\textbf{A}_L(t) = \frac{1}{2\pi}\int_{-\infty}^\infty r^{-1}(\omega)\tilde{\textbf{p}}_L(\omega)e^{i\omega t}d\omega$, $r^{-1}(\omega)$ decreases exponentially with increasing $\omega$. Designating $\omega_c$ as a critical frequency when $r^{-1}(\omega_c)$ is sufficiently small. To ensure that no relevant frequency components are filtered out in the retrieval process, a requirement of $\delta t$ therefore is:
\begin{eqnarray}
\label{eqn:rule1}
\omega_{Nyquist} = \frac{2\pi}{2 \delta t} \ge \omega_c
\end{eqnarray}
In our numerical demonstrations with the proposed test laser fields, a timestep of about $\delta t \sim 2~a.u.$ is used; with this timestep, $|r^{-1}(\omega_{Nyquist})| \sim 10^{-10}$. $\delta t$ can be increased in realistic experiments. With this $\delta t$, the Nyquist $\omega_{Nyquist}$ corresponds to a wavelength of about $29nm$, in the deep ultraviolet regime. This high frequency component (if with substantial intensity) will strongly distorts the tunneling ionization process, rendering our method invalid. Therefore $\delta t$ can be further increased.

%\emph{Optimal filtering} may be needed if some random noises are introduced when measuring the peak shift $\textbf{p}_L(\Delta \tau)$ corresponding to the GIM\cite{press2007numerical}.
\begin{figure}
    \includegraphics[width=0.55\linewidth]{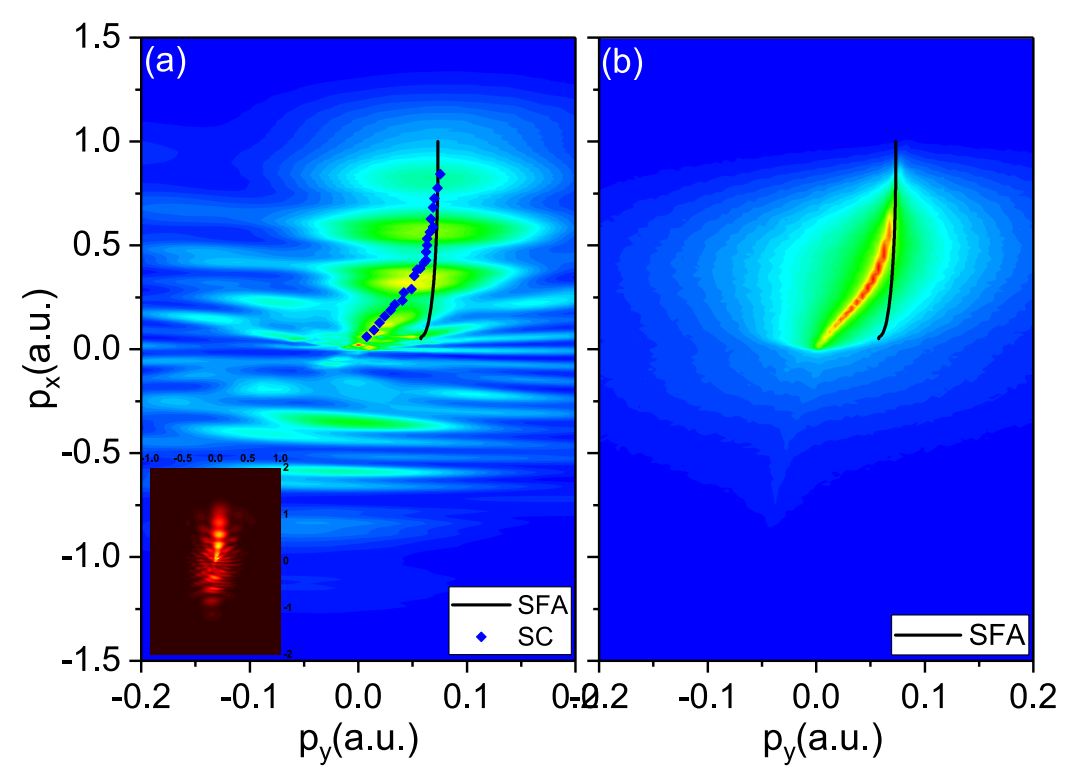}
    \caption{\label{fig:IntegratedPMD} Integrated photoelectron momentum distribution(PMD) in the polarization direction, ionization of H atom by an OTC laser field with time delay $\Delta \tau = 0$. TDSE and SC Monte Carlo simulations for (a) and (b) respectively. Inset(a): Holographic interference pattern by TDSE simulation. Blue diamonds in panel (a) are the peak shift calculated by SC in panel (b).  }
\end{figure}

Aside from the restriction on the timestep $\delta t$, the smallest transverse momentum $\delta p_\perp$ increment that can be distinguished by two consecutive measurement steps is also restricted. A rough estimation can be obtained from Eqn.~\ref{eqn:peak_shift_SFA}, assuming $E_L(t) \approx \epsilon_L \cos\omega_L t,~t_r-t_0 \ll \frac{2\pi}{\omega_L}(t_0=\mathcal{R}e(t_0^R))$:
\begin{eqnarray}
\label{eqn:rule2}
\delta p_\perp  = | \mathcal{R}e\{ \frac{1}{t_r - t_0^R} \int^{t_r}_{t_0^R}\textbf{E}_L(t+\Delta \tau)dt\}| \delta t \lesssim \epsilon_L \frac{\sinh\omega_L t_i}{\omega_L t_i} \delta t
\end{eqnarray}
where $t_i = \mathcal{I}m(t_0^R)$. In the recent study on photon momentum partition and nondipole effects in strong field ionization\cite{PhysRevLett.106.193002, PhysRevLett.113.243001, Hartung2019}, experimentalists are able to resolve the photoelectron transverse momenta amounting to a few photons, on the order of $\delta p_c \sim 10^{-3}~a.u.$.  With $\delta t$ used in our simulation, the upper bounds of $\delta p_\perp$ with our test light pulses(Fig.~3 and Fig.~4) are about $0.005~a.u.$ and $0.01~a.u.$. Test laser pulses with larger intensities can be used in experiments\cite{PhysRevLett.121.253203, PhysRevLett.122.183202}. One may need to resort to statistical spectral analysis for a more detailed estimation.
%and $\delta p_c$ is the experimental resolution of the transverse photoelectron momenta

Concluding this section, the electric field of the test laser field is directly expressed as:
\begin{eqnarray}
\label{eqn:efield}
\textbf{E}_L(t) = \frac{1}{i\pi}\int_{-\infty}^\infty \frac{\omega \tilde{\textbf{p}}_L(\omega)}{a(\omega)+a^\ast(-\omega)} e^{i \omega t}d\omega
\end{eqnarray}

\section{Semiclassical(SC) trajectory Monte Carlo simulation}

In our SC model\cite{liujie2014}, the electron tunnels through the potential
barrier formed by Coulomb potential of the atomic core and the instantaneous
laser electric field. The photoelectron (released at the tunneling exit $r_0$ from the ion) has a Gaussian transverse (with respect to the
direction of instantaneous electric field) velocity distribution: $f(v_\perp) = \exp[-\kappa v^2_\perp/|\epsilon(t_0)|]$, $\kappa = \sqrt{2I_p}$ and $\epsilon(t)$ is the instantaneous laser electric field strength\cite{PhysRevA.87.041403, Delone:91}. Subsequently, the electron moves in combined
laser electromagnetic field and Coulomb potential governed by
Newton's equations of motion: $\frac{d\textbf{p}}{dt} = -(\textbf{E}_0(t)+\textbf{E}_L(t)) - \frac{\textbf{r}}{r^3}$. A large ensemble of electron trajectories on the order of $10^6$ is simulated for analysis.

Fig.~\ref{fig:IntegratedPMD} illustrates the integrated two-dimensional PMD: TDSE and SC trajectory Monte Carlo simulations for (a)(b) respectively. The laser parameters are the same as in Fig.~1.

According to the Coulomb glory rescattering theory\cite{PhysRevLett.121.143201}, the transverse momentum distribution in the near-forward direction for a preassigned $p_x$ would behave like:
\begin{eqnarray}
f(p_\perp) \sim J_0^2(b_g \sqrt{(p_y - p_{y, L})^2+(p_z-p_{z, L})^2})
\end{eqnarray}
The GIM in two independent directions($y$ and $z$) can be extracted from the final 3D momentum distribution simultaneously, without interfering from the other direction.

\begin{figure}
    \includegraphics[width=0.45\linewidth]{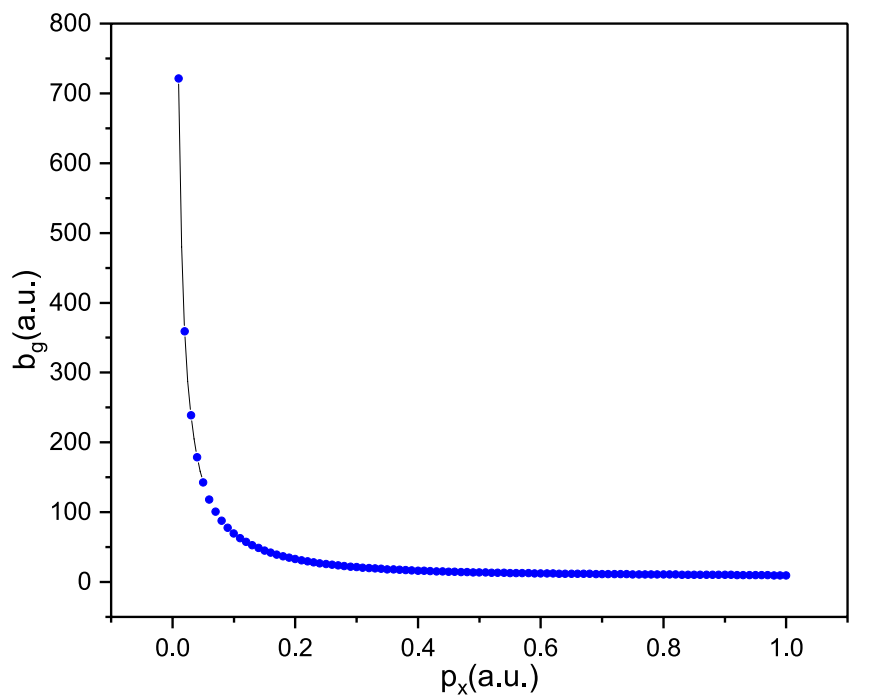}
    \caption{\label{fig:impact_factor} Extracted asymptotic impact factor of the glory trajectories for different longitudinal momentum $p_x$.}
\end{figure}

Using a similar shooting method with a preassigned final
longitudinal momentum $p_x$ as in reference\cite{PhysRevLett.121.143201}, we can reversely extract the initial conditions for the glory trajectory,
thus getting the value of the asymptotic impact factor $b_g$ for every $p_x$. See Fig.~\ref{fig:impact_factor}.

A pure SC trajectory Monte Carlo simulation would yield a logarithmic-like peak structure for the transverse momentum distribution(black dotted lines in Fig.~1(b1)(b2)); this results from the saddle points in the classical deflection function(G1, G3 in Fig.~1(c) are two saddle points of $p_y = p_y(\textbf{p}_{\perp 0})|_{\eta_0=0.3}$)\cite{PhysRevLett.108.033201}. Therefore, if the saddle points or, equivalently, the circle contour corresponding to the GTs are not in the regime of significance of the initial transverse momentum distribution, the Coulomb glory rescattering effect would diminish. Back to the case discussed in our research, this would roughly require that the electron drift due to the weak test laser should not be too large; a very rough estimation would be:
\begin{eqnarray}
\label{eqn:rule3}
\frac{\epsilon_L}{\omega_L} \lesssim \sqrt{\frac{\epsilon_0}{\kappa}}
\end{eqnarray}
The same criterion for an elliptically-polarized strong laser field has already been found\cite{Dan_k_2018}. Moreover, in nondipole strong field ionization, the transverse momentum drift due to the radiation pressure scales as $U_p/c$, where $U_p=\frac{\epsilon_0^2}{4\omega_0^2}$ is the ponderomotive potential and $c$ is the speed of light in vacuum. Since the same Coulomb glory rescattering effect causes the counterintuitive peak shift of the transverse momentum distribution in laser propagation direction\cite{PhysRevLett.113.243001, PhysRevA.95.011402, PhysRevA.100.023413, Hartung2019}, similar criterion $\frac{U_p}{c} \lesssim \sqrt{\frac{\epsilon_0}{\kappa}}$ holds\cite{PhysRevLett.118.093001}. The momentum drift due to a true electric field will typically be much larger than that due to the radiation pressure $\frac{\epsilon_L}{\omega_L} \gg \frac{U_p}{c}$, so the peak position of transverse momentum distribution may be easier to measure in experiments for our method. For the laser parameters and timestep chosen in our simulations, $U_p/c$ is on the same order of $\delta p_\perp$. Fortunately for a particular $p_x$, the peak shift due to radiation pressure of the fundamental laser field is a fixed value\cite{PhysRevA.100.023413, Hartung2019}, therefore if the test laser field is superposed with its polarization parallel to the propagation direction of the fundamental laser field, then a constant value due to the radiation pressure should be subtracted from the measured peak shift of the transverse momentum distribution.

\begin{figure}
    \includegraphics[width=0.6\linewidth]{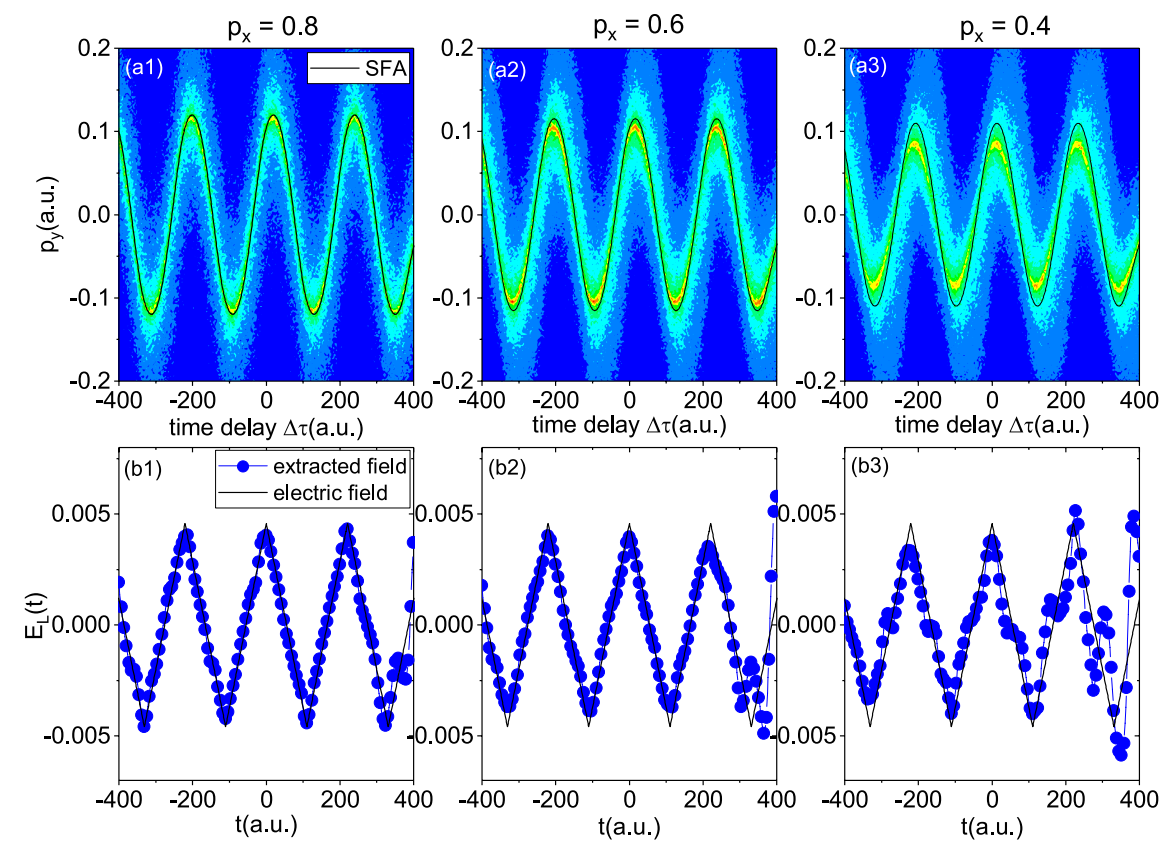}
    \caption{\label{fig:triangularWave}(a1-a3)Streaked photoelectron transverse momentum versus time delay for a triangle test light pulse, the reconstructed electric field is shown in panels(b1)(b2)(b3).}
\end{figure}

Now further inspecting the FRF(Fig.~\ref{fig:FRF}), the frequency components near a particular $\omega_a \approx 0.11~a.u.$ (corresponding to about $400~nm$) are the mostly attenuated. Analyses indicate that $\omega_a \approx 2\omega_0$. Therefore for our scheme to encompass shorter wavelengths, we can decrease the wavelength of the fundamental laser field to, e.g. $400~nm$. Generally we intend our method to measure the waveforms in the visible, infrared or even terahertz regimes, with relatively longer wavelengths. From Fig.~\ref{fig:FRF}, we also conclude that if the test light pulses only contain components with wavelengths longer than about $\lambda_0=800nm$, the FRF can be approximated as:
\begin{eqnarray}
\label{eqn:approximateFRF}
r(\omega) \approx (1- t^2_\beta \omega^2)e^{i(\mp\pi+t_\alpha \omega)}
\end{eqnarray}
 $t_{\alpha, \beta} = t_{\alpha, \beta}(t_0^R,t_r)$ are small time parameters that are determined by the fundamental ionizing laser field(for larger $p_x$ when the Coulomb influence is small), and decrease with increasing $\omega_0$. In the current setup, $t_\alpha \approx 35~a.u.,~t_\beta \approx 9~a.u$. Then we have the approximate relation:
  \begin{eqnarray}
  \textbf{p}_L(\Delta \tau) \approx -\textbf{A}_L(\Delta \tau + t_\alpha) + t_\beta^2 \frac{d\textbf{E}_L}{d\Delta \tau}(\Delta \tau + t_\alpha)
   \end{eqnarray}
   the second term on the right hand side is much smaller than the first term in the current setup. Numerical simulations confirm this relationship. If we \emph{naively} decrease the wavelength of the fundamental ionizing field indefinitely, the same result as that in attosecond streak camera will be achieved. Of course this is not possible for our scheme.

 The reason is that since the fundamental laser field $\textbf{E}_0$ is used to initiate the strong
field tunneling ionization process, its intensity and wavelength should be adjusted to fall into
the tunneling ionization regime. A rough requirement would be that the Keldysh
parameter should be small\cite{keldysh1965ionization}: $\gamma = \sqrt{\frac{I_p}{2U_p}} \lesssim 1$, $U_p = \frac{\epsilon_0^2}{4\omega^2_0}$
is the ponderomotive
potential. Calculations with different combinations of the intensities and carrier frequencies of the fundamental ionizing laser fields indicate that Eqn.~\ref{eqn:approximateFRF} is a very good approximation for the FRF $r(\omega)$ for when $\omega \lesssim \omega_0$.
A choice of the ionizing target with a smaller $I_p$ may by advantageous. In experiments, rare gas atoms like krypton or xenon may be used. Increasing the intensity of the ionizing fundamental laser field may also be considered, but be careful of the saturation of strong field ionization. Some of these restrictions(Eqn.~\ref{eqn:rule1}, Eqn.~\ref{eqn:rule2}, and Eqn.~\ref{eqn:rule3}) may be relaxed a little due to a seemingly redundancy of data, which will be discussed briefly below.

 %Following the last section, we can roughly conclude from the three rough
%estimations(Eqn.~\ref{eqn:rule1}, Eqn.~\ref{eqn:rule2} and Eqn.~\ref{eqn:rule3}) that our theory is more suited to the retrieval of electromagnetic %waveforms with longer wavelengths(the optical spectra of near-infrared, mid-infrared, far-infrared or even terahertz regimes).

\section{Demonstration of the utility of our theory with more waveforms of the test laser fields}

 Fig.~\ref{fig:triangularWave} depicts the calculated results with a triangle wave. The streaking traces are shown in panels (a1)(a2)(a3) for different final longitudinal momentum. In panel (b1)(b2)(b3), the extracted waveform by Eqn.~\ref{eqn:efield} are also illustrated.

\begin{figure}
\includegraphics[width=0.7\linewidth]{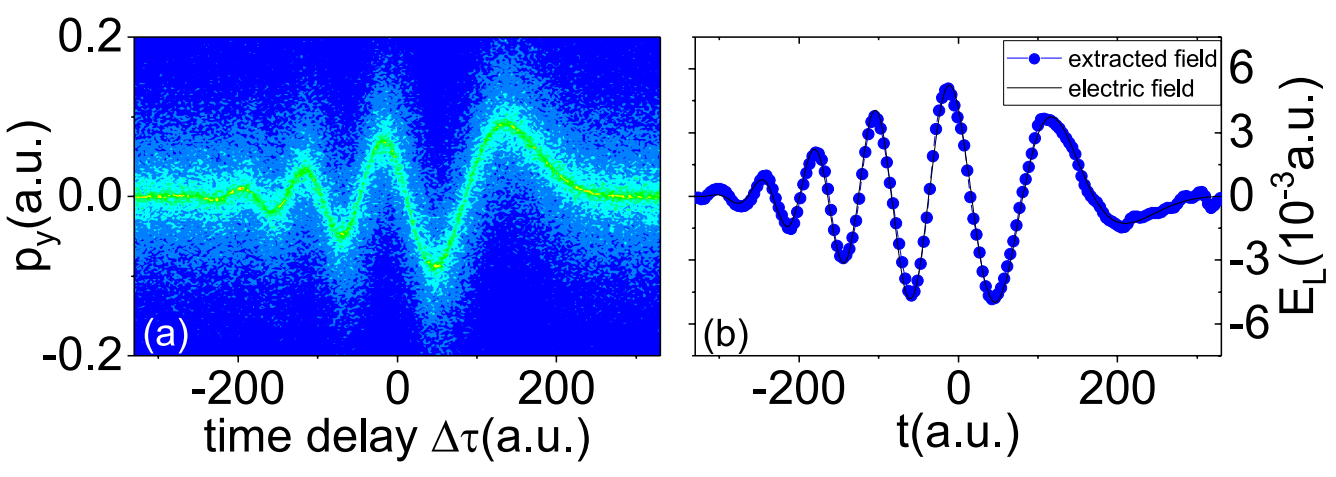}
\caption{\label{fig:linearChirped}(a)Streaking spectra for an octave-spanning linearly chirped light pulse($p_x = 0.8$). (b) The retrieved electric field(blue dotted line).     }
\end{figure}

\begin{figure}
\includegraphics[width=0.7\linewidth]{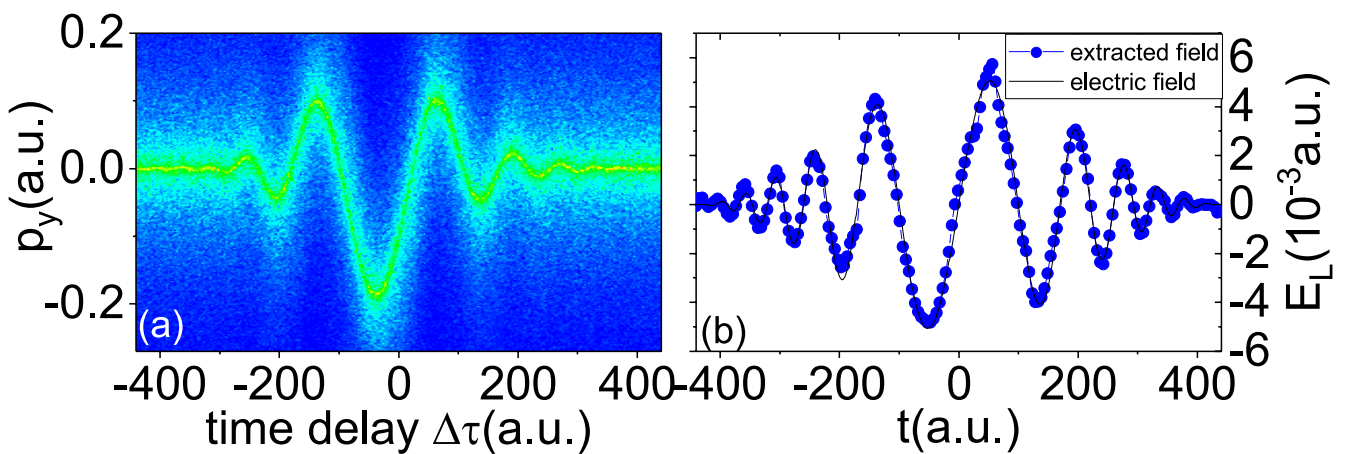}
\caption{\label{fig:quadraticChirped}(a)Streaking spectra for a quadratically chirped light pulse($p_x = 0.8$). (b) The retrieved electric field(blue dotted line).     }
\end{figure}

In Fig.~\ref{fig:linearChirped}(a)(b), the streaking photoelectron transverse momentum spectra and extracted waveform are depicted for an octave-spanning linearly chirped pulse: $E_L(t) = \epsilon_L f_L(t)\cos(\omega_L t - \frac{\omega_L^2}{12\pi}t^2+\psi)$, carrier frequency is $\omega_L = 0.057 ~a.u.$, and time duration $T_L = 6\times \frac{2\pi}{\omega_L}$. $\psi$ is chosen so that $\int E_L(t)dt = 0$. And $\epsilon_L = 0.08 \times \epsilon_0$.

In Fig.~\ref{fig:quadraticChirped}(a)(b), the streaking photoelectron transverse momentum spectra and extracted waveform are depicted for an quadratically chirped pulse: $E_L(t) = \epsilon_L f_L(t)\sin(\omega_L t + \frac{\omega_L^3}{8\pi^2}t^3)$, carrier frequency is $\omega_L = 0.028 ~a.u.$, and time duration $T_L = 4\times \frac{2\pi}{\omega_L}$. And $\epsilon_L = 0.08 \times \epsilon_0$.

\section{Discussion of the possibility of a new streak camera scheme in the attosecond/subfemtosecond regime}

In principle, our method is very analogous to the attosecond angular streaking (attoclock) technique\cite{Eckle1525, Eckle2008, Sainadh2019}. While in attoclock, the tunneling photoelectron  wavepacket(TPW) released by the strong near-circularly-polarized laser field is streaked to different angles by the same field, yielding an one-to-one mapping between the tunneling time and the final direction of the asymptotic electron momenta. In our scheme, the TPW is released by a strong linearly-polarized laser field, careful tailoring of the waveform of this pump pulse may lead to a more confined TPW in the time domain\cite{Wirth195}.  Unfortunately, the glory trajectories released at different tunneling time all contribute to the same direction(parallel to the fundamental laser polarization). Introducing a perpendicularly-polarized weak probe field lifts this degeneracy, yielding the one-to-one correspondence between the tunneling time and the direction of infinite glory trajectories(Fig. 1(a), for different $p_x$(different tunneling time), the glory interference maximum is different): $t_0 \mapsto \textbf{p}_L$.

\begin{figure}
\includegraphics[width=0.6\linewidth]{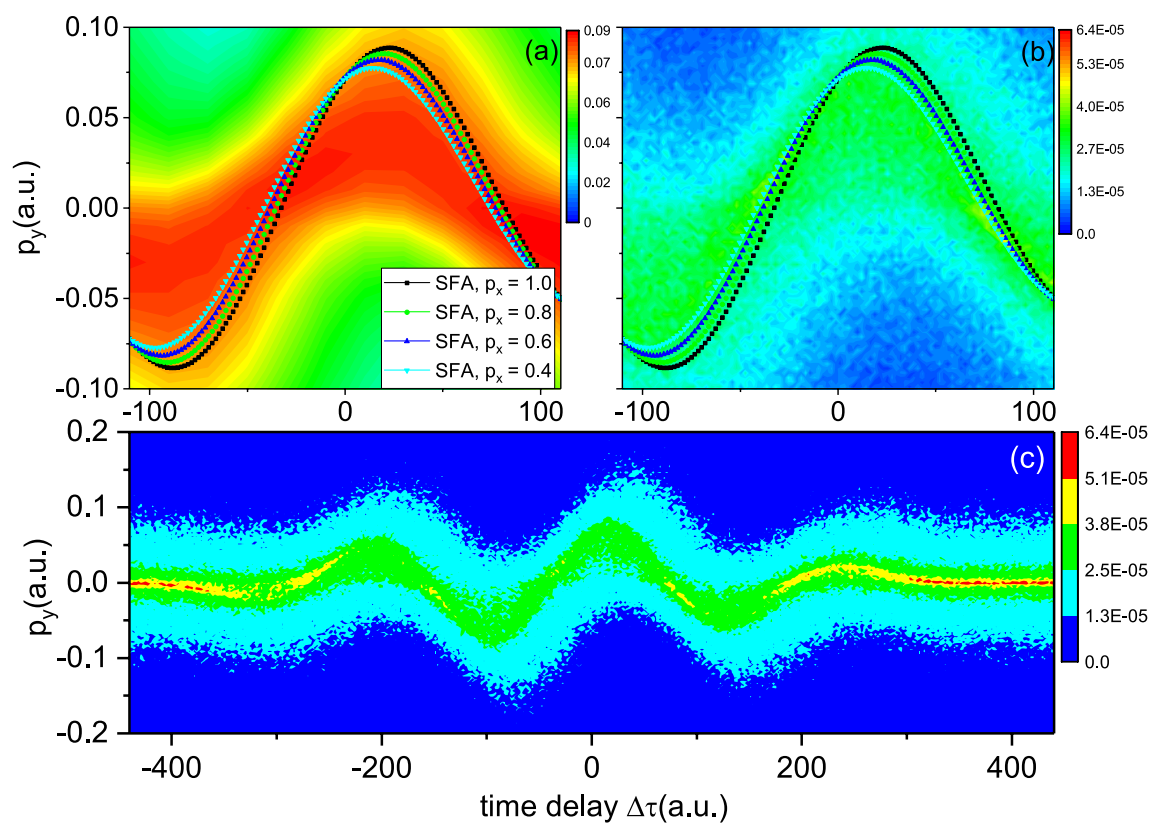}
\caption{\label{fig:integratedStreakingTrace} Integrated streaking asymptotic momentum $p_y$ spectra versus time delay between the pump(central wavelength $800nm$, peak intensity $1.5 \times 10^{14} W/cm^2$, time duration three optical cycles) and probe(central wavelength $1600nm$, peak intensity $2.4 \times 10^{11} W/cm^2$, time duration four optical cycles) lights. (a): TDSE calculation. (b)(c): SC trajectory Monte Carlo simulation results. In panels(a)(b), dotted lines are calculated by SFA with fixed longitudinal momenta $p_x$.}
\end{figure}

Moreover, our scheme has an extra control knob compared to attoclock(in which the strong near-circularly-polarized laser field acts both as the pump and the probe): the time delay between the fundamental and test laser pulses can be varied. This indicates that our method also bears some similarities to the attosecond streak camera; in both cases, the photoelectron transverse momentum/energy distribution for every time delay is a mapped replica of some properties of the initial photoelectron transients\cite{Kienberger2004}. In Fig.~\ref{fig:integratedStreakingTrace} the integrated streaking transverse momentum spectra versus time delay are demonstrated(with the longitudinal momenta $p_x$ integrated out), the streaking traces are obviously more "broadband"(along the $p_y$ axis). The "bandwidth" is a measure of the TPW in the time domain.

This seemingly redundancy of data may lead to much more robust retrieval schemes. In attosecond streak camera, this is reflected in the fact that with a time-delayed streaking spectrum, both the temporal structures of the attosecond XUV and near-infrared light pulses can be retrieved double-blindly with accuracy, and the algorithm is very robust against noises\cite{Wang_2009}. Our scheme may finally evolve to a similar status. However, before that, a feasible theory that can nicely account for the Coulomb effects should be developed(Eqn.~\ref{eqn:peak_shift_SFA} fails especially for small $p_x$ due to Coulomb potential influence). Although this theory is still lacking, our present work represents a first step toward a new streaking scheme in the attosecond/subfemtosecond regime.

\bibliographystyle{apsrev4-1}
%\bibliography{references}
\input{supplementary.bbl}

%% file: SubfsGloryHologrammetry.bbl
%merlin.mbs apsrev4-1.bst 2010-07-25 4.21a (PWD, AO, DPC) hacked
%Control: key (0)
%Control: author (72) initials jnrlst
%Control: editor formatted (1) identically to author
%Control: production of article title (-1) disabled
%Control: page (0) single
%Control: year (1) truncated
%Control: production of eprint (0) enabled
%

%% file: supplementary.bbl
%merlin.mbs apsrev4-1.bst 2010-07-25 4.21a (PWD, AO, DPC) hacked
%Control: key (0)
%Control: author (72) initials jnrlst
%Control: editor formatted (1) identically to author
%Control: production of article title (-1) disabled
%Control: page (0) single
%Control: year (1) truncated
%Control: production of eprint (0) enabled
%